\documentclass[aps,pre,amsmath,amssymb,floatfix,nofootinbib,showkeys, preprint]{revtex4-2}
\usepackage{mathtools}
\begin{document}
\title{The origin of the Poisson distribution in stochastic dynamics of gene expression}
\author{Julian Lee}
\email{jul@ssu.ac.kr}
\affiliation{Department of Bioinformatics and Life Science, Soongsil University, Seoul, Korea}
\date{\today}
\begin{abstract}
The Poisson distribution is the probability distribution of the number of independent events in a given period of time. Although the Poisson distribution appears ubiquitously in various stochastic dynamics of gene expression, both as time-dependent distributions and the stationary distributions, underlying independent events that give rise to such distributions have not been clear, especially in the presence of the degradation of gene products, which is not a Poisson process. I show that, in fact, the variable that follows the Poisson distribution is the number of independent events where biomolecules are created, which are destined to survive until the end of a given time duration. This new viewpoint allows us to derive time-dependent Poisson distributions as solutions of master equations for general class of protein production and degradation dynamics, including models with time-dependent rates and a non-Markovian model with delayed degradation. I then derive  analytic forms of general  time-dependent probability distributions by combining the Poisson distribution with the binomial or the multinomial distributions.
\end{abstract}
\keywords{Gene Regulatory Network, Stochastic fluctuation, Poisson distribution}
\maketitle
\newpage
\section{Introduction}
Gene regulatory network(GRN) controls the life process by producing and degrading various kinds of proteins that perform important biological functions. It is a well-known fact that the time evolution of mRNA and/or protein molecules in such a network is stochastic~\cite{that,elo,swain,paul,rase,gold05,newman06,cai,fried,yu06,lipshtat,raj08,elo2,tani10,delay1,delay2,neu13,gold13,kumar,hao2,sung,lee18,ham20,lin21}. Even when various external conditions such as cellular environments are identical, there is always  intrinsic noise due to remaining uncontrolled factors that influence the GRN of interest, making its dynamics stochastic. It has been suggested that biological organisms may have evolved to take advantage of such fluctuations~\cite{elo2}. 

In simple theoretical models of stochastic GRN dynamics, the Poisson distribution,
\begin{equation}
P_{\rm Poisson}(n;\mu) = e^{-\mu} \frac{\mu^n}{n!}, \label{poi}
\end{equation}
often appears as a probability distribution for the number of mRNA or protein molecules, both as time-dependent and stationary distributions~\cite{that,swain,delay1,delay2,sung,ham20,lin21}. In fact, the Poisson distribution arises from a Poisson process, where the probability of an event occurring during a short time interval $[t,t+dt]$ is independent of events in other time regions~\cite{kampen2,stbook,stbook2}. Now, consider a simple transcription process where a gene is always active, and an mRNA molecule $X$ is transcribed with a rate $\alpha$, 
\begin{equation}
\varnothing \xrightarrow{\alpha}  X. \label{birth1}    
\end{equation}
It is clear that the creation events are indeed independent events, forming a Poisson process. Therefore, if we start from zero molecule at $t=0$, then the number $n$ of mRNA molecules at any later time $t$ is exactly the same as the number of creation events up to that time-point, and consequently, the Poisson distribution is actually the distribution of the number of creation events during the time interval $[0,t]$.  However, consider a model where a degradation 
\begin{equation}
X \xrightarrow{\beta}  \varnothing.    
\end{equation}
is also included. Now, the degradation 
is {\it not} a Poisson process, although sometimes it is erroneously written as such in the literature. Since a molecule that already got degraded cannot be degraded again, the probability of a degradation event happening during the short time period $[t,t+dt]$ definitely depends on how many degradation events happened in times earlier than $t$. However, the Poisson distributions still  appear ubiquitously in this class of models as the distributions of the number of mRNA or protein molecules~\cite{that,swain,delay1,delay2,sung,ham20,lin21}, making its origin mysterious. Therefore, I address the following question in this paper: Given that the Poisson distribution appears so ubiquitously in the stochastic dynamics of GRN as the distribution of molecule number of mRNA or protein, is this molecule number equal to the number of certain independent events that happened during a given time interval? As I will show, the answer to this question is affirmative. In fact, it is the number of events where mRNA molecules are created, which are destined to be survive until the end of a given interval. The answer is very simple once stated, almost on the verge of being trivial.  The most probable reason that it has not been discussed in the literature is that it may have been  counter-intuitive to consider a birth of a particle with a given fate. However, it is important to note that although the fate of a molecule is not determined at the time of its creation, the {\it probability} of its given fate at the end of the time interval, death or survival, is already determined at the time of its creation, and it is all that matters in defining a Poisson process. 

Using this viewpoint of Poisson processes, I can not only rederive Poisson distribution for models of gene expression with time-dependent rates and a non-Markovian model with time delay, I can also derive analytic forms of general time-dependent distributions by combining the Poisson distribution with other distributions. In fact,  time-dependent distributions for {\it arbitrary} initial distributions are obtained for the Markovian model by combining the Poisson distribution with the binomial distribution, and  distributions for very general class of initial distributions are obtained for a non-Markovian model by combining the Poisson distribution with the multinomial distribution.

The remainder of the paper is organized according to the order of increasing  complexity. In section II, I will briefly review Poisson processes, where I will emphasize the importance of inhomogeneous Poisson processes, consisting of independent events that do not necessarily follow identical distributions. In section III, I will consider molecule creations without degradations, a textbook example of a  Poisson process. In section IV, I will describe the molecule degradation, which is ${\it not}$ a Poisson process. I show that a general time-dependent probability distribution can be expressed as a superposition of binomial distributions.  In section V, I will describe molecule creations with degradations and the underlying Poisson process. I show that any time-dependent distribution can be expressed in terms of the Poisson and the binomial distributions. The result will be generalized to a non-Markovian model with delayed degradation in section VI, where I will not only rederive the Poisson distribution in an earlier work with less effort, but also derive a general time-dependent distribution  by combining the Poisson distribution with the multinomial distribution.  Finally, the discussions are given in section VII.

\section{The Poisson distribution and Poisson Processes}
The Poisson distribution in Eq.(\ref{poi}) is specified by a single parameter $\mu$, the expected number of independent events in a given time duration.  
 If we partition the time interval $[0,t]$ into $N$  sub-intervals of small size $\Delta t \equiv t/N$, then 
 the probability of events happening more than once in each sub-interval is $O\left((\Delta t)^2\right)$, and the probability of single occurrence of the event takes the form $p=\lambda \Delta t + O\left((\Delta t)^2\right)$ when such a probability is time-independent. Therefore, the total number of events in $[0,t]$ approximately follows the binomial distribution\footnote{The range of $n$ in the probability  distribution $P(n,t)$ will be considered to be all the integers without any restriction in this work, unless specified otherwise.  This is acceptable as long as we set $P(n,t)=0$ for illegitimate values of $n$. All the explicit forms of probability distributions considered here such as binomial, multinomial, and Poisson distributions contain factorials of  negative integers in the denominators whenever the value of the molecule number is out of legitimate range, and   vanish due to the fact that $(j!)^{-1} = [\Gamma(j+1)]^{-1}=0$ whenever $j$ is a negative integer. Taking the range of the molecule number to be the whole integer is especially convenient for summation, where the summation index can be shifted freely, which I will do throughout the text without further elaboration.},
 \begin{equation}
P_{\rm binom}(n;\{ N,p\}) =  \frac{N!}{n!(N-n)!}p^n (1-p)^{N-n},  \label{bin}
\end{equation}
the distribution of the number of successes in $N$ independent and identical trials, with success probability of $p$ at each trial. 
 The Poisson distribution is recovered by taking the limit of $N \to \infty$ with $\mu(t) = Np = \lambda t$ fixed~(Appendix \ref{binpoi}): 
\begin{equation}
P_{\rm Poisson}(n; \mu(t)) =  \frac{e^{-\mu(t)}}{n!}\mu(t)^n,  \label{poi2}
\end{equation}
where $\mu(t)$ is the expected number of events in the time-interval $[0,t]$.

An important point to note here is that the condition of identical trials is not actually required to obtain the Poisson distribution. We only have to require {\it independent} trials~\cite{kampen2,stbook}. In this case, the binomial distribution in Eq.(\ref{bin}) is generalized to
\begin{equation}
\tilde P_{\rm \ binom}(n; \{p_1, \cdots, p_N\}) =  \sum_{\{i_1 < i_2  \cdots < i_n\}} p_{i_1} p_{i_2} \cdots p_{i_n} \prod_{k \notin \{i_1 , \cdots , i_n\}} \left(1-p_k \right)   \label{bin2}
\end{equation}
where the probability of success at $i$-th trial is $p_i$, and the summation is over all distinct set of $n$ indices $\{i_1<i_2, \cdots < i_n\} \subset \{1,2, \cdots N\}$. 
Again,  the distribution in Eq.(\ref{bin2}) becomes the Poisson distribution in the limit of $\Delta t \to 0$,  now with the time-dependent function $\lambda(t)$, so that $\mu(t) = \int_0^t \lambda(t')$~\cite{stbook}(Appendix \ref{binpoi}). This is the process where the probability of an event happening in an infinitesimal time interval $[t,t+dt]$ is $\lambda(t) dt$, called an inhomogeneous Poisson process, to distinguish it from the case with constant $\lambda$,  called a homogeneous Poisson process. 

\section{Molecule creation without degradation}
This is a simple birth process
\begin{equation}
\varnothing \xrightarrow{\alpha(t)}  
 X,\label{birth}    
\end{equation}
where $X$ describes an mRNA or protein molecule, and the creation rate $\alpha(t)$ is time-dependent in general\footnote{All the time-dependent rates in this work are the rates with predetermined time-dependences, and the corresponding model should not be confused with the one where the rates themselves are stochastic variables~\cite{tele,sung,ham20,lin21}.}.  Since molecule creations at distinct time points are independent of each other, these creation events form a Poisson process. If the number of molecules is zero at $t=0$, then the molecule number $n$ at a later time $t>0$ is equal to the number of creation events in the time interval $[0,t]$. Therefore, the probability distribution of $n$ is given by the Poisson distribution, $P(n,t) = P_{\rm Poisson}(n;\mu(t))$, with $\mu(t) = \int^t_0 \alpha(t') dt'$.

Even when the number of molecules takes a non-zero value $n_0$ at $t=0$, the number of molecules $n$ at a later time $t>0$ is simply $n_0+n'$ where $n'$ is the number of creation events in the time interval $[0,t]$. Therefore, $n'=n-n_0$ still follows the Poisson distribution, leading to a shifted Poisson distribution for $n$,
\begin{equation}
P(n,t) = \frac{e^{-\mu(t)}}{(n-n_0)!} \mu(t)^{n-n_0}. \label{shift}
\end{equation}

The dynamics of the probability distribution is Markovian, described by the master equation
\begin{equation}
\frac{\partial P(n,t)}{\partial t} = \alpha(t) \left[ P(n-1,t) - P(n,t) \right].   \label{master}
\end{equation}
It is straightforward to check that Eq.(\ref{shift}) is a solution of the master equation (\ref{master}) by direct substitution. By the linearity of the master equation (\ref{master}), an analytic form of the general solution with an arbitrary initial distribution $P(n,0) = v(n)$ can be constructed by superposing the expressions in  Eq.(\ref{shift}),
\begin{equation}
P(n,t) = \sum_{n_0}  \frac{e^{-\mu(t)}}{(n-n_0)!} \mu(t)^{n-n_0} v(n_0).  \label{gen}
\end{equation}
From Eq.(\ref{gen}), we see that the master equation (\ref{master}) admits  stationary solutions
\begin{equation}
    P_{\rm st}(n) \sum_{n_0}  \frac{e^{-\mu(\infty)}}{(n-n_0)!} \mu(\infty)^{n-n_0} v(n_0)
\end{equation}
that depend on initial distributions, 
if and only if $\mu(\infty) = \int_0^\infty \alpha(s) ds$ exists. In particular, for a constant rate $\alpha$, there is no stationary solution because $\mu(t)=\alpha t$ increases indefinitely with time, and $P(n,t)$ converges to zero pointwise for all $n$. In this case, all the states in the Markov chain are transient states~\cite{markov}. 

\section{Model only with degradation}
This is a simple death process
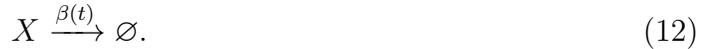
\begin{equation}
X \xrightarrow{\beta(t)}  \varnothing.\label{death}    
\end{equation}
In contrast to the creation process in the previous section, this process is {\it not} a Poisson process, because any molecule that has been degraded cannot be degraded again. Therefore, the degradation events are not independent, and the probability of a degradation event  depends on the number of molecules available. The master equation describing the dynamics of this model is
\begin{equation}
\frac{\partial P(n,t)}{\partial t} = \beta(t) \left[ (n+1) P(n+1,t) - n P(n,t) \right].   \label{master2}
\end{equation}
 In order to get a non-zero number of molecules, the initial number $n_0$ of  molecules must be non-zero. The number of molecules at later times can be obtained by using the fact that the event of a given molecule surviving at a later time $t$ is independent of what happens to other molecules, whose probability $p_{\rm surv}(t,0)$ is given by
\begin{equation}
    p_{\rm surv}(t,0) = \exp\left(-\int_0^t \beta(s) ds\right). \label{surprob}
\end{equation}
Therefore, the number $n$ of surviving molecules at time $t$ follows the binomial distribution
   \begin{equation}
    P(n,t) = P_{\rm binom}(n;\{ n_0,p_{\rm surv} (t,0) \}) = \frac{n_0!}{n!(n_0-n)!} p_{\rm surv} (t,0)^n (1-p_{\rm surv} (t,0))^{n_0-n}. \label{binsol}
\end{equation} 
It can be checked that the expression in Eq.(\ref{binsol}) is the solution of the master equation (\ref{master2}) by direct substitution. Again, the general solution with an arbitrary initial distribution $P(n,0) = v(n)$ can be constructed by the superposition
\begin{equation}
    P(n,t) = \sum_{n_0} \frac{n_0!}{n!(n_0-n)!} p_{\rm surv} (t,0)^n (1-p_{\rm surv} (t,0))^{n_0-n} v(n_0). \label{gen2}
\end{equation} 
A Poisson distribution appears in the special case where the initial distribution is Poissonian:
\begin{equation}
    v(n) = \frac{e^{-\mu_0}\mu_0^n}{n!}, 
\end{equation}
in which case we obtain the Poisson distribution $P(n,t) = P_{\rm Poisson}(n;\mu(t))$ with
\begin{equation}
    \mu(t) = \mu_0 p_{\rm surv}(t,0), 
\end{equation}
due to the fact that the superposition of binomial distributions weighted by a Poisson distribution is again a Poisson distribution~\cite{stbook}(Appendix \ref{poipoi}). In the next section, I will show that this distribution can be interpreted as a result of a Poisson process.

When the integral $\int_0^t \beta(s) ds$ diverges as $t \to \infty$, as in the case of a constant $\beta$, we get  $\lim_{t \to \infty} p_{\rm surv} (t,0) = 0$. In this case, we see from Eq.(\ref{gen2}) that the probability distribution approaches the stationary distribution
\begin{equation}
   \lim_{t \to \infty} P(n,t) = P_{\rm st}(n) = \delta_{n,0},
\end{equation}
regardless of the initial distribution.  

\section{Model with both creation and degradation}
This model is described by
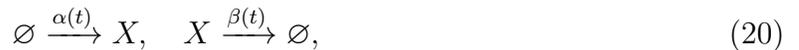
\begin{equation}
\varnothing \xrightarrow{\alpha(t)}  X,\quad
X \xrightarrow{\beta(t)}  \varnothing, \label{bd}
\end{equation}
which belongs to a class of models called the birth-death models~\cite{kampen2,stbook,stbook2}. Let us first consider the case where the number of $X$ molecules is initially zero. Molecules are created in the time interval $[0,t]$, but only a fraction of them survive at $t$, which I will call surviving molecules. When a molecule is created in the short time interval $[t',t'+dt']$ with $t'<t$, its fate at $t$ is  undetermined, but the probability of its survival $p_{\rm surv}(t,t')$ at time $t$ is already determined to be
\begin{equation}
    p_{\rm surv}(t,t')= \exp\left(-\int_{t'}^t \beta(s)ds\right),\label{psurv2}
\end{equation}
and consequently, the probability that a surviving molecule is created in  $[t',t'+ dt']$ is given by $\alpha(t')p_{\rm surv}(t,t')dt'$, independent of events that happen in other regions of time. Therefore, the number of molecules at a later time $t$ is equal to the number these independent events in $[0,t]$, following the Poisson distribution $P_{\rm Poisson}(n;\mu(t))$ with 
\begin{equation}
    \mu(t) = \int_0^t \alpha(s) p_{\rm surv}(t,s)ds= \int_0^t \alpha(s) \exp\left(-\int_{s}^t \beta(u)du\right)ds. \label{genmu}
\end{equation}
The master equation for this model is
\begin{equation}
\frac{\partial P(n,t)}{\partial t} = \alpha(t)\left[  P(n-1,t) -  P(n,t) \right] + \beta(t) \left[ (n+1) P(n+1,t) - n P(n,t) \right],   \label{master3}
\end{equation}
and it is straightforward to check by direct substitution that $P_{\rm Poisson}(n;\mu(t))$ with $\mu(t)$ given by Eq.(\ref{genmu}) is a solution of this equation. 

We noted in the previous section that the distribution of the molecule number in the degradation-only model is Poissonian for $t>0$ if the initial distribution at $t=0$ is also Poissonian. In fact, this distribution can be considered as a special case of the Poisson distribution in the model with a time-dependent creation rate $\alpha(t)$, by shifting the origin of time: We start from zero molecule at some time $t_0<0$ so that we obtain a Poisson distribution for $P(n,0)$. We then require that $\alpha(t)=0$ for $t>0$ so that the model reduces to the degradation-only model for $t>0$.

The probability distribution converges to a stationary Poisson distribution if and only if the expected number of surviving molecules at $t \to \infty$,  given by the integral 
\begin{equation}
    \mu(\infty) = \int_0^\infty \alpha(s) \exp\left(-\int_{s}^\infty \beta(u)du\right)ds, \label{genmust}
\end{equation}
 is finite. For example, if the rates $\alpha$ and $\beta$ are constants, we get
\begin{equation}
    \mu(t) = \int_0^t \alpha e^{-(t-s) \beta}ds = \frac{\alpha}{\beta}(1-e^{-\beta t}),  \label{conmu}
\end{equation}
and
\begin{equation}
    \mu(\infty) =  \frac{\alpha}{\beta}.  \label{conmust}
\end{equation}

Now consider a more general situation where the initial number of molecules is $n_0$. The number of molecules at a later time $t$ is the sum of the number $n_1$ of surviving molecules among initial $n_0$ particles, whose distribution follows the binomial distribution in Eq.(\ref{binsol}), and the number $n_2$ of surviving molecules created during the interval $[0,t]$, which follows the Poisson distribution. Therefore, the probability distribution for the molecule number $n$ is
\begin{eqnarray}
    P(n,t) &=& \sum_{n_1+n_2=n} P_{\rm binom}(n_1;\{ n_0,p_{\rm surv} (t,0) \}) P_{\rm Poisson}(n_2; \mu(t)) \nonumber\\
    &=& \sum_{n_1} \frac{n_0!}{n_1! (n_0-n_1)!}  \frac{e^{-\mu(t)} \mu(t)^{n-n_1}}{(n-n_1)!} p_{\rm surv} (t,0)^{n_1} (1-p_{\rm surv} (t,0))^{n_0-n_1} \label{gensol}
\end{eqnarray}
where $p_{\rm surv} (t,0)$ and $\mu(t)$ are given by Eq.(\ref{surprob}) and Eq.(\ref{genmu}), respectively. It is straightforward to check that the solution (\ref{gensol}) is a solution of the master equation (\ref{master3}) by direct substitution~(Appendix \ref{binpoisol}). By the linearity of the master equation, the solution for an  arbitrary initial distribution $P(n,0)=v(n)$ can be constructed by superposing the distributions in Eq.(\ref{gensol}),
\begin{eqnarray}
    P(n,t) 
    &=& \sum_{n_0,n_1} \frac{v(n_0) n_0! e^{-\mu(t)} \mu(t)^{n-n_1}}{n_1! (n_0-n_1)!   (n-n_1)!} p_{\rm surv} (t,0)^{n_1} (1-p_{\rm surv} (t,0))^{n_0-n_1}.
    \label{gensol2}
\end{eqnarray}
\section{Model with time-delayed degradation}
Due to a complex mechanism of protein degradation, there can be a time-delay in protein degradation.  Model with time-delayed degradation attempts to capture such a behavior, and once a molecule enters the degradation process, it gets degraded after a fixed time delay $\tau$~\cite{delay1}. A more general model where the molecule is allowed to get degraded before $\tau$ was also constructed~\cite{delay2}, which I will examine in this section in detail. The model is
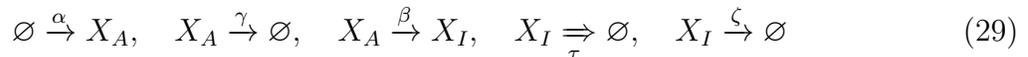
\begin{equation} \varnothing \xrightarrow{\alpha}  X_A,\quad 
X_A \xrightarrow{\gamma}\varnothing,\quad
X_A \xrightarrow{\beta} X_I,\quad
X_I \xRightarrow[\tau]{} \varnothing,\quad
X_I \xrightarrow{\zeta}\varnothing
\label{dbd}
\end{equation}
Here, $X_I$ is an inactive molecule that has entered the degradation process. An active molecule is denoted by $X_A$, which can undergo instantaneous degradation with rate $\gamma$. It can enter delayed degradation process with rate $\beta$, at which point it becomes inactive and get degraded after time $\tau$ with certainty, but it can also undergo instantaneous degradation with rate $\zeta$ before the delayed degradation process is complete.  
Although it is straightforward to allow for time-dependent rates, I will keep them time-independent in order to compare the result with that of ref.~\cite{delay2}, as well as for notational simplicity. 

This model is non-Markovian, and the master equation for the probability distribution $P(n_A,n_I,t)$ takes the form
\begin{eqnarray}
\frac{d P(n_A,n_I,t)}{dt} &=& (E_A^{-1} -1) \alpha P(n_A,n_I,t) + (E_A -1) \gamma n_A  P(n_A,n_I,t) \nonumber\\
&+& (E_A E_I ^{-1} -1) \beta n_A  P(n_A,n_I,t)  + (E_I -1) \zeta n_I P(n_A,n_I,t)\nonumber\\
&+& \sum_{n'_A} P^*(n_A,n_I-1,\tau| n'_A -1) \beta n'_A P(n'_A, t-\tau) e^{-\zeta \tau},\label{nmm} 
\end{eqnarray}
where $E^k_{A,I} f (n_{A,I} ) \equiv f (n_{A,I} + k)$, $P(n_A, t)\equiv \sum_{n_I} P(n_A,n_I,t)$ is the marginal probability for $n_A$, and $P^*(n_A,n_I,t|n'_A-1)$ is the  probability under the initial condition of $n'_A-1$ active molecules and no inactive molecules, obtained by neglecting the time-delayed degradation~\cite{delay2}:
\begin{eqnarray}
\frac{d P^*(n_A,n_I,t|n'_A-1)}{dt} &=& (E_A^{-1} -1) \alpha P^*(n_A,n_I,t|n'_A-1) + (E_A -1) \gamma n_A  P^*(n_A,n_I,t|n'_A-1) \nonumber\\
&+& (E_A E_I ^{-1} -1) \beta n_A  P^*(n_A,n_I,t| n'_A -1) \nonumber\\
&+& (E_I -1) \zeta n_I P^*(n_A,n_I,t| n'_A -1). \label{eqcond}
\end{eqnarray}
  Eq. (\ref{nmm}) along with  Eq. (\ref{eqcond}) admits the Poisson distribution as a time-dependent solution if the particle number is initially zero or the initial distribution is Poissonian~\cite{delay2}. However, it is easy to derive the Poisson distribution without going through the complicated process of solving Eqs.(\ref{nmm}) and (\ref{eqcond}), by considering the underlying Poisson process.

Note that one of three things happen in a given short time interval of $[t',t'+dt']$: either a molecule that will survive at $t$ as an active one is created, the one that will be survive as an inactive one is created\footnote{This should not be confused with a birth of an inactive molecule. Every molecule is born active, and only an active molecule can turn into an inactive one. Here I am considering a birth of an active molecule that will eventually survive as an inactive molecule at the later time $t$}, or none of these happens. These events are independent of creation events in other regions in time. Also, although the creation of a surviving active molecule and a surviving inactive molecule are exclusive events, they can be treated as independent events because they are rare events: even if we assume they are independent, probability of more than two molecules created in $[t',t'+dt']$ is negligible. Therefore, exclusive and independent events are indistinguishable if they are rare~(Appendix \ref{mulpoi}), and  the probability distribution for the numbers $n_A$, $n_I$ of $X_A$, $X_I$ are given by the product of Poisson distributions,
\begin{equation}
    P(n_A,n_I, t) = \frac{e^{-\mu_A(t)-\mu_I(t)}}{n_A! n_I!} \mu_A(t)^{n_A} \mu_I(t)^{n_I}.\label{poi31}
\end{equation}
It only remains to compute $\mu_A(t)$ and $\mu_I(t)$.  

Suppose that a molecule is created at time $t'$. In order for the molecule to remain active at a later time $t$, (i) it should not undergo instantaneous degradation ($X_A \xrightarrow{\gamma}\varnothing$), and (ii) should not turn into an inactive molecule $(X_A \xrightarrow{\beta} X_I)$, during the intervening time. Therefore, the conditional probability $p_A(t-t')$ that this molecule will remain active at $t$ is given by the  exponential distribution in $t-t'$,
\begin{equation}
    p_A(t-t')= e^{-a (t-t')}, \label{pAsurv}
\end{equation}
where $a\equiv \beta + \gamma$,
and we get
\begin{equation}
    \mu_A(t) = \int_{0}^t \alpha p_A(t-t')d t' = \frac{\alpha}{a}  \left(1 -  e^{-a t}\right). \label{mua}
\end{equation}
For a molecule created at time $t'$ to survive at a later $t$ as an inactive molecule, (i) it must turn into an inactive one within a short time interval $[t_I,t_I + dt_I]$ such that $\max (t',t-\tau) < t_I < t$, (ii) it should not undergo instantaneous degradation ($X_A \xrightarrow{\gamma}\varnothing$) in the time between $t'$ and $t_I$, and (iii) it should not undergo instantaneous degradation ($X_I \xrightarrow{\zeta}\varnothing$) in the time between $t_I$ and $t$. For a given value of $t_I$, such a conditional probability $\rho_I(t,t_I/t')dt_I$ is given as
\begin{equation}
    \rho_I(t,t_I/t')dt_I = e^{-a (t_I-t')} e^{-\zeta (t-t_I)}\beta dt_I 
\end{equation}
Integrating over $t_I$, we get the conditional probability $p_I(t-t')$ that a molecule created at time $t'$ will survive as an inactive one at $t$:
\begin{eqnarray}
    p_I(t-t') &=& \int_{\max (t',t-\tau)}^t \rho_I(t,t_I/t')d t_I =   \frac{\beta e^{a t'-\zeta t}} {\zeta-a}  \left( e^{(\zeta -a)t}- e^{(\zeta -a){\max(t',t-\tau)}} \right)    \nonumber\\
    &=& \left\{
   \begin{array}{ll}
        {\beta} (\zeta-a)^{-1}   \left(e^{a(t'-t)} - e^{\zeta(t'-t)} \right)     & \quad (0 \leq t - t' \leq \tau),\\
       {\beta} (\zeta-a)^{-1}  e^{a(t'-t)} \left(1 - e^{(a - \zeta) \tau} \right)    & \quad  (\tau \leq t- t' < t).
    \end{array} \right.\label{pIsurv}
    \end{eqnarray}
The expected number of surviving inactive molecules, $\mu_I(t)$, is then obtained by multiplying $p_I(t-t')$ by the creation rate $\alpha$ and integrating over $t'$: 
\begin{eqnarray}
    \mu_I(t) = \int_{0}^t \alpha p_I(t-t')d t' &=& \left\{
   \begin{array}{ll}
        \frac{\alpha \beta} {\zeta-a}  \left[\frac{1}{a}(1 - e^{- a t}) -\frac{1}{\zeta }(1 - e^{-\zeta t})\right]  & \quad  (0 \leq t < \tau),\\
       \frac{\alpha \beta} {a}  \left[\frac{1 - e^{-\zeta\tau}}{\zeta} +  \frac{1 - e^{( a-\zeta)\tau}}{ a-\zeta}e^{- a t} \right]     & \quad (t \geq \tau ).
    \end{array} \right.\label{mui}
\end{eqnarray}
Eq.(\ref{poi31}) along with Eq.(\ref{mua}) and Eq.(\ref{mui}) completely specify the time-dependent Poisson distribution when we start from zero molecule at $t=0$, which agrees  with the result in ref.~\cite{delay2}.  The total number of molecules,  $n=n_A + n_I$, also follows a Poisson distribution with the expectation value $\mu(t) = \mu_A(t) + \mu_I(t)$, due to the fact that the sum of two variables that follow Poisson distributions also follows a Poisson distribution~\cite{stbook2}. The Poisson process underlying the Poisson distribution for $n=n_A + n_I$ is the creation of particles  destined to survive until time $t$, regardless of being active or inactive.

    
    Finally, as in the case of the Markovian model in the previous section, we can derive general time-dependent distributions by combining the Poisson distribution with the multinomial distribution. However, note that it is almost impossible to allow arbitrary initial distribution at $t=0$. If a non-zero initial value of $n_I$ is allowed, there is an ambiguity in the evolution of the system for $t>0$ because the final degradation of the initial inactive molecules depend on the exact time points at $t<0$ that they got inactivated. Therefore, I will only consider the case where only active particles are present at $t=0$, whose number is $n^0$. Given an active molecule at $t=0$, the probability that it will survive as an active molecule and the probability that it will survive as an inactive molecule, at a later time $t$, are given by Eqs.(\ref{pAsurv}) and (\ref{pIsurv}), respectively, with $t'=0$:
    \begin{eqnarray}
        p_A(t) &=& e^{-a  t},\nonumber\\
        p_I(t) &=& \left\{
   \begin{array}{ll}
        {\beta} (\zeta-a)^{-1}   \left(e^{- a t} - e^{-\zeta t} \right)    & \quad  (t < \tau),\\
         {\beta} (\zeta-a)^{-1}  e^{- a t} \left(1 - e^{(a - \zeta) \tau} \right)   & \quad (t \geq \tau).
    \end{array} \right.\label{pApI}
    \end{eqnarray}
    Therefore, the probability that there are $n'_{A}$ surviving active molecules and $n'_{I}$ surviving active molecules {\it among} initial $n_0$ active molecules, at time $t$, is given by the multinomial distribution,
    \begin{eqnarray}
        P_{\rm mult}(n'_{A},n'_{I};\{n_0,p_A(t),p_I(t)\}) &\equiv& \frac{n_0!}{n'_{A}! n'_{I}! (n_0-n'_{A}-n'_{I})!} p_A(t)^{n'_{A}}p_I(t)^{n'_{I}}\nonumber\\
        &&\times (1-p_A(t)-p_I(t))^{n_0-n'_{A}-n'_{I}}.
    \end{eqnarray}
    The numbers of active and inactive molecules at $t$ are decomposed as $n_A=n'_{A} + n''_{A}$, $n_I=n'_{I} + n''_{I}$, where $n''_{A}$and $n''_{I}$ are the numbers of the surviving active and inactive molecules created in the time interval $[0,t]$.  Therefore, the probability distribution for $n_A$ and $n_I$ at $t$ is given by the combination of the multinomial and the Poisson distribution:
    \begin{eqnarray}
        P(n_A,n_I,t) &=& \sum_{n'_{A}+n''_{A}=n_A} \sum_{n'_{I}+n''_{I}=n_I} P_{\rm mult}(n'_{A},n'_{I};\{n_0,p_A(t),p_I(t)\})\nonumber\\
        &\times& P_{\rm Poisson}(n''_{A}; \mu_A(t)) P_{\rm Poisson}(n''_{I}; \mu_I(t)) \nonumber\\
        &=& \sum_{n'_{A}} \sum_{n'_{I}} \frac{n_0!}{n'_{A}! n'_{I}! (n_0-n'_{A}-n'_{I})!} p_A(t)^{n'_{A}}p_I(t)^{n'_{I}}\nonumber\\
        &&\times (1-p_A(t)-p_I(t))^{n_0-n'_{A}-n'_{I}}\nonumber\\
        &&\times \frac{e^{-\mu_A(t)-\mu_I(t)} \mu_A(t)^{n_A-n'_{A}}\mu_I(t)^{n_I-n'_{I}}}{(n_A-n'_{A})!(n_I-n'_{I})!}.\label{mulpoisol1}
    \end{eqnarray}
    We can check that the distribution in Eq.(\ref{mulpoisol1}) is indeed a solution of the master equation Eq.(\ref{nmm}) by direct substitution~(Appendix \ref{mulpoisol}). 
    Again, by the linearity of the master equation, the time-dependent probability distribution for an arbitrary initial distribution of {\it active} molecules,
    \begin{equation}
        P(n^0_A,n^0_I,0) = v(n^0_A) \delta_{n^0_I,0},
    \end{equation}
    is obtained by the superposition of Eq.(\ref{mulpoisol1}) weighted by $v(n_0)$,
    \begin{eqnarray}
        P(n_A,n_I,t) 
        &=& \sum_{n_0} \sum_{n'_{A}} \sum_{n'_{I}} v(n_0) \frac{n_0!}{n'_{A}! n'_{I}! (n_0-n'_{A}-n'_{I})!} p_A(t)^{n'_{A}}p_I(t)^{n'_{I}}\nonumber\\
        &&\times (1-p_A(t)-p_I(t))^{n_0-n'_{A}-n'_{I}}\nonumber\\
        &&\times \frac{e^{-\mu_A(t)-\mu_I(t)} \mu_A(t)^{n_A-n'_{A}}\mu_I(t)^{n_I-n'_{I}}}{(n_A-n'_{A})!(n_I-n'_{I})!}. 
    \end{eqnarray}
    
\section {Discussions}
Poisson distributions appear  ubiquitously in stochastic dynamics of  gene expression, and the Poisson noise is considered to be the most basic type of noise when analyzing various components of stochastic fluctuations.  However, when gene products are allowed to get degraded, it has not been clear whether the molecule number following a Poisson distribution is equal to the  number of certain independent events in time, and if this is the case, what is the corresponding events.  I answered this question in this work, by showing that the number of molecules distributed according to the Poisson distribution is actually equal to the number of creations of the molecules that are destined to survive until the end of a given time period, which are indeed independent events in time that form an inhomogeneous Poisson process. Using this viewpoint, I could derive the Poisson distribution not only for the  Markovian model with time-dependent-rates, but also for the model with time-delayed degradation,  without performing the difficult task of solving the non-Markovian master equation. Furthermore, I could obtain the time-dependent distribution for an  arbitrary initial distribution in the case of the Markovian model, by combining the Poisson distribution with  the binomial distribution, and the time-dependent distribution for an arbitrary initial distribution of {\it active} molecules in the case of the non-Markovian model, by combining the Poisson distribution with the multinomial distribution.

Of course, the molecule number follows the Poisson distribution only under the  simplifying assumption of independent creation events. In the case of the protein, the number of protein molecules follows the Poisson distribution only if we approximate the transcription and the translation as a one-step process. In reality, the mRNA molecule has a non-zero life-time, during which protein gets translated with a certain rate, leading to bursty translations~\cite{cai,fried,yu06}. The creations of protein molecules are not a Poisson process in such a model, since the translation rate depends on the mRNA concentration. The  Poissonian description also breaks down in most of the models with stochastic rates~\cite{tele,sung,ham20,lin21}. In this class of models, the creation rate itself is a random variable distributed according to a probability distribution. A model with a stochastic creation rate can be used as an approximate model to describe the regulation of the expression by the transcription factor binding~\cite{tele,ham20}, or to emulate the extrinsic noise due to heterogeneous cellular environments~\cite{ham20}. When the creation rate is a stochastic variable, the creations of the molecules form a Poisson process only if the creation rate has no temporal correlations, which is not valid for most of the non-trivial models. Despite these limitations, general analytic solutions found in this work may be of value as a basis of perturbations to obtain more realistic descriptions of the gene-regulatory network. For example, one may numerically solve the master equation for the stochastic rate, and then use the analytic solution for a {\it given} realization of the creation rate, so that the expectation values of various physical quantities are expressed as weighted averages over  numerical distributions of the stochastic rates. Similar analyses may be done for other sophisticated models of gene regulatory network by combining the analytic solutions found in the current work with other analytic solutions and/or numerical computations.

\begin{acknowledgments} 
This work was supported by the National Research Foundation of Korea, funded by the Ministry of Science and ICT (NRF-2020R1A2C1005956). 
\end{acknowledgments}
\bibliography{stoch2}

\appendix
\section{The Poisson distribution as the limit of the distribution for independent trials}\label{binpoi}
 The number of successes in $N$ independent trials, with the success probability at $i$-th trial being $p_i$, follows the probability distribution which is a generalization of the binomial distribution, 
\begin{equation}
\tilde P_{\rm \ binom}(n; \{p_1, \cdots, p_N\}) =  \sum_{\{i_1 < i_2  \cdots < i_n\}} p_{i_1} p_{i_2} \cdots p_{i_n} \prod_{k \notin \{i_1 , \cdots , i_n\}} \left(1-p_k \right)   \label{abin2}
\end{equation}
 I want to show that the Poisson distribution can be obtained from this expression in the limit of $N\to \infty$ with $\mu=\sum_{j=0}^N p_j$ fixed.  One can derive the desired result using the generating function $F(z) \equiv \sum_n P(n) z^n$. On one hand, the generating function $\tilde F(z; \{p_1, \cdots, p_N\})_{\rm binom}$ for the distribution $\tilde P_{\rm \ binom}(n; \{p_1, \cdots, p_N\})$  is
\begin{eqnarray}
\tilde F(z; \{p_1, \cdots, p_N\})_{\rm binom} &\equiv& \sum z^n \tilde  P_{\rm \ binom}(n; \{p_1, \cdots, p_N\}) \nonumber\\
&=& \sum_{n=0}^N z^n  \sum_{\{i_1 < i_2  \cdots < i_n\}} p_{i_1} p_{i_2} \cdots p_{i_n} \prod_{k \notin \{i_1 , \cdots , i_n\}} \left(1-p_k \right)\nonumber\\
&=&  \sum_{n=0}^N  \sum_{\{i_1 < i_2  \cdots < i_n\}} (zp_{i_1})( z p_{i_2}) \cdots (z p_{i_n}) \prod_{k \notin \{i_1 , \cdots , i_n\}} \left(1-p_k \right)\nonumber\\
&=&\prod_{j=1}^N \left[z p_j + (1-p_j)\right] \label{bingen}
\end{eqnarray}
On the other hand, the generating function $F_{\rm Poisson}(z)$ for the Poisson distribution is
\begin{equation}
    F_{\rm Poisson}(z; \mu) = \sum_n \frac{z^n \mu^n e^{-\mu}}{n!} = e^{\mu(z-1)}. \label{poigen}
\end{equation}
Now, Eq.(\ref{bingen}) can be rewritten as
\begin{eqnarray}
\tilde F(z; \{p_1, \cdots, p_N\})_{\rm binom} &=& \prod_{j=1}^N \left[1 + (z - 1)  p_j \right]=  \prod_{j=1}^N \left[ e^{(z - 1)  p_j} + O(p_j^2)  \right]\nonumber\\
&=& e^{\mu(z-1)} + O(1/N).\label{bingen3}
\end{eqnarray}
Therefore,
\begin{equation}
\lim_{N \to \infty} \tilde F(z; \{p_1, \cdots, p_N\})_{\rm binom} = F_{\rm Poisson}(z; \mu),
\end{equation}
from which we deduce
\begin{equation}
\lim_{N \to \infty} \tilde P_{\rm \ binom}(n; \{p_1, \cdots, p_N\}) = P_{\rm \ Poisson}(n; \mu).
\end{equation}

Since the binomial distribution is a special case of $\tilde P_{\rm \ binom}(n; \{p_1, \cdots, p_N\})$ with $p_j=p$ for all $j$, it is easy so see that the Poisson distribution is obtained from the binomial distribution by taking the limit of $N \to \infty$ with $\mu=Np$ fixed.

\section{The superposition of binomial distribution weighted by a Poisson distribution is again a Poisson distribution}\label{poipoi}
Consider a binomial distribution for $n_0$ independent and identical trials with success probability $p$ at each trial, and suppose that $n_0$ is itself stochastic, distributed with Poisson distribution with the expectation value  $\mu$. Then the number $n$ of success again follows a Poisson distribution, with the expected number events being $\mu p$~\cite{stbook}:
\begin{equation}
    \sum_{n_0} P_{\rm Poisson}(n_0; \mu) P_{\rm binom}(n; \{ n_0,p \}) = P_{\rm Poisson}(n; \mu p). \label{popo}
\end{equation}
The Poisson distribution at the left-hand side of Eq.(\ref{popo}) can be considered to come from a Poisson process where an event  happens within a short time interval $[t,t+dt]$ with probability $\lambda(t) dt$ , so that $\mu=\int_0^t \lambda (t') dt'$. Now, whenever such an event happens, we also toss a coin with head probability $p$, and count the event only when the coin produces the head. It is intuitively clear that the number of counted events in the time interval $[0,t]$ obviously follows the Poisson distribution with expected number $\mu p$, as given in the right-hand side of Eq.(\ref{popo}). Before formally proving Eq.(\ref{popo}), I first prove the discrete version obtained by replacing the Poisson distributions in Eq.(\ref{popo})  by the binomial distributions,
\begin{equation}
    \sum_{n_0} P_{\rm binom}(n_0; \{N, q \}) P_{\rm binom}(n; \{ n_0,p \}) = P_{\rm binom}(n; \{N, qp \} ),\label{twocoin}
\end{equation}
which may be easier to grasp intuitively. This expression arises in the situation where we toss two coins $N$ times,  the probability of the head for two coins at each trial being $p$ and $q$, respectively. The success for a given trial is defined as the event that both coins produce heads. Then the number of successes follows the binomial distribution with the success probability $pq$ at each trial, which is the right-hand side of Eq.(\ref{twocoin}). The left-hand side is its decomposition using a conditional probability. We first compute the probability $P_{\rm binom}(n_0; \{N, q \})$ that the first coin produced heads $n_0$ times. We then compute the probability $P_{\rm binom}(n; \{ n_0,p \})$ that {\it among} $n_0$ trials with the first coin producing heads, $n$ of them have the second coin producing the heads.  It is intuitively clear that the summation at the left-hand side is equal to the right-hand side, but one can also  explicitly show that
\begin{eqnarray}
&&\sum_{n_0} P_{\rm binom}(n_0; \{N, q \}) P_{\rm binom}(n; \{ n_0,p \}) \nonumber\\
&=& \sum_{n_0} \frac{N!}{n_0! (N-n_0)!}q^{n_0} (1-q)^{N-n_0} \times \frac{n_0!}{n!(n_0-n)!} p^n (1-p)^{n_0-n} \nonumber\\
&=& \frac{N!}{(N-n)!n!}(pq)^n\sum_{n_0} \frac{(N-n)!}{ (N-n_0)!(n_0-n)!}   q^{n_0-n} (1-p)^{n_0-n} (1-q)^{N-n_0} \nonumber\\
&=& \frac{N!}{(N-n)!n!}(pq)^n\sum_{j} \frac{(N-n)!}{ (N-n-j)!j!}   [q (1-p)]^{j}(1-q)^{N-n-j} \nonumber\\
&=& \frac{N!}{(N-n)!n!}(pq)^n\left[ q(1-p) + 1-q \right]^{N-n} = \frac{N!}{(N-n)!n!}(pq)^n\left[ 1-pq \right]^{N-n} \nonumber\\
&=& P_{\rm binom}(n; \{N, qp \}),
\end{eqnarray}
proving Eq.(\ref{twocoin}). It is straightforward to extend the proof to the inhomogeneous case where $p$ and $q$ are different for each trial, and Eq.(\ref{popo}) is then obtained in the limit of $N \to \infty$ with $\mu=\sum_{i=1}^N q_i$ fixed. However, one can also prove Eq.(\ref{popo})  directly:
\begin{eqnarray}
&&\sum_{n_0} P_{\rm Poisson}(n_0; \mu) P_{\rm binom}(n; \{ n_0,p \})\nonumber\\
&=& \sum_{n_0} \frac{e^{-\mu} \mu^{n_0}}{n_0!} \times \frac{n_0!}{n!(n_0-n)!} p^n (1-p)^{n_0-n} = \frac{e^{-\mu}}{n!}(\mu p)^n\sum_{n_0}  \frac{\mu^{n_0-n}}{(n_0-n)!}  (1-p)^{n_0-n} \nonumber\\
&=& \frac{e^{-\mu}}{n!}(\mu p)^n\sum_{j}  \frac{[\mu (1-p)]^{j}}{j!} =  \frac{e^{-\mu}}{n!}(\mu p)^n e^{\mu (1-p)} = \frac{e^{-\mu p}}{n!}(\mu p)^n = P_{\rm Poisson}(n; \mu p). 
\end{eqnarray}
     
\section{The  time-dependent distribution of the Markovian model with non-zero initial number of molecules (Eq.(\ref{gensol})) is the solution of the master equation (\ref{master3}).}\label{binpoisol}     
To show that the distribution given by Eq.(\ref{gensol}) is the solution of the master equation, it is convenient use the generating function $F(z,t)\equiv \sum_n P(n,t) z^n$. Then the master equation (\ref{master3}) turns into
\begin{equation}
    \partial_t F(z,t) = (z-1)(\alpha(t) -\beta(t)\partial_z) F(z,t). \label{genmaster}
\end{equation}
The generating function for the distribution in Eq.(\ref{gensol}) is
\begin{eqnarray}
    F(z,t) &=& \sum_n z^n \sum_{n_1} \frac{n_0!}{n_1! (n_0-n_1)!}  \frac{e^{-\mu(t)} \mu(t)^{n-n_1}}{(n-n_1)!} p_{\rm surv} (t,0)^{n_1} (1-p_{\rm surv} (t,0))^{n_0-n_1} \nonumber\\
    &=& \sum_{n_1} z^{n_1}  \frac{n_0!}{n_1! (n_0-n_1)!}p_{\rm surv} (t,0)^{n_1} (1-p_{\rm surv} (t,0))^{n_0-n_1}  \times \sum_{n_2} \frac{e^{-\mu(t)} \mu(t)^{n_2}z^{n_2}}{(n_2)!}  \nonumber\\
    &=& \left[1 + p_{\rm surv} (t,0) (z-1)\right]^{n_0} e^{\mu(t)(z-1)}
\end{eqnarray}
On one hand, by substituting $F(z)$ to the left-hand side of Eq.(\ref{genmaster}), we get
\begin{eqnarray}
    \partial_t F(z,t) &=& n_0 (z-1) \left[1 + p_{\rm surv} (t,0) \right]^{n_0-1} \dot p_{\rm surv} (t,0) e^{\mu(t)(z-1)}\nonumber\\
    &+&  (z-1) \left[1 + p_{\rm surv} (t,0) (z-1)\right]^{n_0} e^{\mu(t)(z-1)} \dot \mu(t).\label{oh}
\end{eqnarray}
On the other hand, by substituting $F(z)$ to the right-hand side of Eq.(\ref{genmaster}), we get
\begin{eqnarray}
    (z-1)(\alpha(t)-\beta(t) \partial_z) F(z,t) &=& \alpha(t) (z-1) \left[1 + p_{\rm surv} (t,0) (z-1)\right]^{n_0}   e^{\mu(t)(z-1)}\nonumber\\
    &-& n_0 \beta(t) (z-1)  \left[1 + p_{\rm surv} (t,0) (z-1)\right]^{n_0-1}  p_{\rm surv} (t,0) e^{\mu(t)(z-1)}\nonumber\\
    &-& \beta(t) (z-1) \mu(t) \left[1 + p_{\rm surv} (t,0) (z-1)\right]^{n_0} e^{\mu(t)(z-1)} \label{oth}
\end{eqnarray}  
Since
\begin{eqnarray}
    \dot p_{\rm surv}(t,0) &=& -\beta(t) p_{\rm surv}(t,0),\nonumber\\
    \dot \mu(t) &=& \alpha(t)  -\beta(t) \mu(t),
\end{eqnarray}
which can be easily checked by taking the time derivatives of $p_{\rm surv} (t,0)$ and $\mu(t)$ in Eqs.(\ref{psurv2}) and (\ref{genmu}),
we see that the expressions in Eq.(\ref{oh}) and Eq.(\ref{oth}) are equal. Therefore, the distribution in Eq.(\ref{gensol}) is the solution of the master equation (\ref{master3}).

\section{The Poisson distribution is the limit of the distribution for independent trials, when there are multiple alternatives }\label{mulpoi}
The binomial distribution arises in independent and identical trials when there are only two alternatives at each trial. When there are multiple alternatives, whose number is $m$,  then the numbers of outcomes $( n_1, \cdots n_{m-1} )$ in the total $N$ trials follow the  multinomial distribution
\begin{equation}
    P_{\rm mult}(n_1, \cdots, n_{m-1};\{N,p_1, \cdots p_{m-1}\}) = \frac{N!}{n_1! n_2! \cdots n_m!} p_1^{n_1} p_2^{n_2} \cdots p_m^{n_m}
\end{equation}
if the probability of $k$-the alternative happening at each trial is $p_k$, where
$n_m = N - \sum_{i=1}^{m-1} n_i$ and $p_m = 1- \sum_{i=1}^{m-1} p_i$. The corresponding generating function is:
\begin{eqnarray}
    F_{\rm mult}(z_1, z_2 \cdots z_{m-1}) &\equiv& \sum_{n_1, \cdots n_{m-1}}P_{\rm mult}\left(n_1, \cdots, n_{m-1}\right) z_1^{n_1} \cdots z_{m-1}^{n_{m-1}}\nonumber\\
    &=& \sum_{n_1, \cdots n_{m-1}}\frac{N!}{n_1! n_2! \cdots n_m!} p_1^{n_1} p_2^{n_2} \cdots p_m^{n_m} z_1^{n_1} \cdots z_{m-1}^{n_{m-1}}\nonumber\\
    &=& \left[p_m + z_1 p_1 + \cdots + z_{m-1} p_{m-1} \right]^N
\end{eqnarray}
It is straightforward to write down the generating function for the inhomogeneous counterpart,
\begin{eqnarray}
    \tilde F_{\rm mult}(z_1, z_2 \cdots z_{m-1})  &=& \prod_{j=1}^N \left[p_m^{(j)} + z_1 p_1^{(j)} + \cdots + z_{m-1} p_{m-1}^{(j)} \right]\nonumber\\
    &=& \prod_{j=1}^N \left[1 + (z_1 - 1)  p_1^{(j)} + \cdots + (z_{m-1} - 1) p_{m-1}^{(j)} \right] ,
\end{eqnarray}
where $p_k^{(j)}$ denotes the probability of the occurrence of $k$-th alternative happening at $j$-th trial, with $p_m^{(j)} \equiv 1 - \sum_{k=1}^{m-1} p_k^{(j)}$. 

We now take the limit $N \to \infty$, with $\mu_k = \sum_j p^{(j)}_k$ fixed for $1 \leq k \leq m-1$. We get
\begin{eqnarray}
    \tilde F_{\rm mult}(z_1, z_2 \cdots z_{m-1})  
    &=& \prod_{j=1}^N \left[\exp\left((z_1-1) p_1^{(j)}  +  \cdots + (z_{m-1} - 1) p_{m-1}^{(j)}\right)  +  O(1/N^2) \right] \nonumber\\
    &=& \exp\left( (z_1-1) \sum_{j=1}^N p_1^{(j)}  +  \cdots + (z_{m-1} - 1) \sum_{j=1}^N  p_{m-1}^{(j)}\right) + O(1/N) \nonumber\\
    &\xrightarrow[N \to \infty] &\quad  \exp\left((z_1-1) \mu_1 + \cdots + (z_{m-1} -1) \mu_{m-1} \right),
\end{eqnarray}
which is nothing but the generating function for $m-1$ independent Poisson distributions, with expected number of occurrences of $k$-th alternative being $\mu_k$. 

Note that for finite $N$, $k$ events with $k=1, \cdots m-1$ are exclusive events and are therefore not independent. However, even if we assume they are independent, probability of such events occurring more than once becomes negligible in the limit $N \to \infty$, because they are rare events with probability being $O(1/N)$. Therefore, the exclusive and independent events become indistinguishable. Let us illustrate this point with the homogeneous case with $m=3$. The corresponding multinomial distribution describes the case where we toss a three-faced coin, with two heads denoted as $A$ and $B$. There are three possible outcomes at each trial: The  head $A$ with probability $p$, the  head $B$ with probability $q$, or tail with probability $1-p-q$. The probability distribution is given by the multinomial distribution,
\begin{equation}
    P(n_A,n_B)  = \frac{N!}{n_A! n_B! (N-n_A - n_B)!} p^{n_A} q^{n_B} (1-p-q)^{N-n_A - n_B},
\end{equation}
which was already shown above to approach the Poisson distribution 
\begin{equation}
    P(n_A,n_B) = \frac{e^{-\mu_A-\mu_B}}{n_A! n_B!} \mu_A^{n_A} \mu_B^{n_B}.\label{poi3}
\end{equation}
 in the limit of $N \to \infty$ with $\mu_A=N p$ and $\mu_B = N q$ fixed.
Now compare this with the case where we toss two independent two-sided coins denoted as $A$ and $B$ at each trial, which produce heads with probabilities $p$ and $q$, respectively. In contrast to the previous model, outcomes of head $A$ and head $B$ are independent, and the simultaneous heads of $A$ and $B$ are now allowed so that there are four outcomes at each trial. The probability distribution for $n_A$ and $n_B$ is now the product of binomial distributions,
 \begin{equation}
    P(n_A,n_B) = \frac{N!}{n_A! (N- n_A)! } p^{n_A}  (1-p)^{N-n_A }\frac{N!}{n_B! (N - n_B)!}  q^{n_B} (1-q)^{N - n_B}.
\end{equation}
Since each binomial distribution approaches Poisson distribution, we again get Eq.(\ref{poi3}) in the limit of $N \to \infty$ with $\mu_A$ and $\mu_B$ fixed. In this limit, the probability $pq$ of both coins producing heads, being $O(1/N^2)$, becomes negligible. Therefore,  independent events happening in the short time interval of size $O(N^{-1})$ become effectively exclusive, or vice versa, in the limit of $N \to \infty$. This is the reason why multinomial distribution, or its inhomogeneous counterpart, factorizes into independent Poisson distributions in this limit.

\section{The  time-dependent distribution of the non-Markovian model with non-zero initial number of molecules (Eq.(\ref{mulpoisol1})) is the solution of the master equation (\ref{nmm}).}\label{mulpoisol}     
To show that the distribution given by Eq.(\ref{mulpoisol1}) is the solution of the master equation, it is convenient use the generating function $G(z,w,t)\equiv \sum_{n_A} \sum_{n_I} P(n_A,n_I, t) z^{n_A} w^{n_I}$. Then the master equation (\ref{nmm}) turns into~\cite{delay2}
\begin{eqnarray}
    \partial_t G (z,w,t) &=& \left[\gamma (1-z)+ \beta (w-z) \right] \partial_z G + \zeta (1-w)\partial_w G + \alpha(z-1)  G(z,w,t)\nonumber\\
    &+& \beta e^{-\zeta \tau} (1-w) \sum_{n_A} \left[1+ \Phi(z,w,\tau)\right]^{n_A-1}n_A P(n_A,t-\tau)\nonumber\\
    &\times& \exp\left[\alpha \int_0^\tau dt'\Phi(z,w,t') \right]\theta(t-\tau),
    \label{nmgen}
\end{eqnarray}
where
\begin{equation}
    \Phi(z,w,t) \equiv (z-1) p_A(t) + (w-1) p_I(t)
\end{equation} with $p_A(t)$ and $p_I(t)$ given in Eq.(\ref{pApI}). The last term in Eq.(\ref{nmgen}) involving $\Phi(z,w,t) $ describes the process where the active molecule that entered the degradation process at $t-\tau$ gets degraded. Therefore, for the initial condition where $n_I=0$ at $t=0$, this term is absent for $t < \tau$,  implemented here by the step function $\theta(t-\tau)$.

From Eq.(\ref{mulpoisol1}), we get the marginal probability distribution:
\begin{eqnarray}
    P(n_A,t) &=& \sum_{n_I} \sum_{n'_{A}} \sum_{n'_{I}} \frac{n_0!}{n'_{A}! n'_{I}! (n_0-n'_{A}-n'_{I})!} p_A(t)^{n'_{A}}p_I(t)^{n'_{I}}\nonumber\\
        &&\times (1-p_A(t)-p_I(t))^{n_0-n'_{A}-n'_{I}}\nonumber\\
        &&\times \frac{e^{-\mu_A(t)-\mu_I(t)} \mu_A(t)^{n_A-n'_{A}}\mu_I(t)^{n_I-n'_{I}}}{(n_A-n'_{A})!(n_I-n'_{I})!} \nonumber\\
        &=&  \sum_{n'_{A}} \sum_{n'_{I}} \frac{n_0!}{n'_{A}! n'_{I}! (n_0-n'_{A}-n'_{I})!} p_A(t)^{n'_{A}}p_I(t)^{n'_{I}}\nonumber\\
        &&\times (1-p_A(t)-p_I(t))^{n_0-n'_{A}-n'_{I}}\nonumber\\
        &&\times \frac{e^{-\mu_A(t)-\mu_I(t)} \mu_A(t)^{n_A-n'_{A}}}{(n_A-n'_{A})!} \sum_{n_I} \frac{\mu_I(t)^{n_I}}{n_I!} \nonumber\\
        &=&  \sum_{n'_{A}} \sum_{n'_{I}} \frac{n_0!}{n'_{A}! n'_{I}! (n_0-n'_{A}-n'_{I})!} p_A(t)^{n'_{A}}p_I(t)^{n'_{I}}\nonumber\\
        &&\times (1-p_A(t)-p_I(t))^{n_0-n'_{A}-n'_{I}}  \frac{e^{-\mu_A(t)} \mu_A(t)^{n_A-n'_{A}}}{(n_A-n'_{A})!}\nonumber\\
        &=&  \sum_{n'_{A}}  \frac{n_0!}{n'_{A}! (n_0-n'_{A})!} p_A(t)^{n'_{A}} (1-p_A(t))^{n_0-n'_{A}}  \frac{e^{-\mu_A(t)} \mu_A(t)^{n_A-n'_{A}}}{(n_A-n'_{A})!}\nonumber\\
        &=& \sum_{n'_{A}} P_{\rm binom}(n'_{A};\{n_0,p_A(t)\}) P_{\rm Poisson}(n_A-n'_A; \mu_A(t))
        ,\label{condap}
\end{eqnarray}
and the generating function for $t \geq 0$:
\begin{eqnarray}
    G(z,w,t)  &=& \sum_{n_{A}}\sum_{n_{I}}\sum_{n'_{A}} \sum_{n'_{I}} \frac{z^{n_A} w^{n_I} n_0!}{n'_{A}! n'_{I}! (n_0-n'_{A}-n'_{I})!} p_A(t)^{n'_{A}}p_I(t)^{n'_{I}}\nonumber\\
        &&\times (1-p_A(t)-p_I(t))^{n_0-n'_{A}-n'_{I}}\nonumber\\
        &&\times \frac{e^{-\mu_A(t)-\mu_I(t)} \mu_A(t)^{n_A-n'_{A}}\mu_I(t)^{n_I-n'_{I}}}{(n_A-n'_{A})!(n_I-n'_{I})!} \nonumber\\
         &=& \sum_{n'_{A}} \sum_{n'_{I}} \frac{z^{n'_A} w^{n'_I} n_0!}{n'_{A}! n'_{I}! (n_0-n'_{A}-n'_{I})!} p_A(t)^{n'_{A}}p_I(t)^{n'_{I}}\nonumber\\
        &&\times (1-p_A(t)-p_I(t))^{n_0-n'_{A}-n'_{I}}\nonumber\\
        &&\times \sum_{n''_{A}}\sum_{n''_{I}} z^{n''_A} w^{n''_I} \frac{e^{-\mu_A(t)-\mu_I(t)} \mu_A(t)^{n''_{A}}\mu_I(t)^{n''_{I}}}{(n''_{A})!(n''_{I})!} \nonumber\\
    &=& \left[1 + p_A(t) (z-1) + p_I(t) (w-1)\right]^{n_0} e^{\mu_A(t)(z-1)+\mu_I(t)(w-1)}\nonumber\\
    &=& \left[1 + \Phi(z,w,t) \right]^{n_0} e^{\mu_A(t)(z-1)+\mu_I(t)(w-1)}.\label{genap}
\end{eqnarray}

First, we compute the summation in the last term at the  right-hand side of Eq.(\ref{nmgen}) by substituting the expressions for $P(n_A,t)$ and $G(z,w,t)$:
\begin{eqnarray}
 &&\sum_{n_A} \left[1+ \Phi(z,w,\tau)\right]^{n_A-1}n_A P(n_A,t-\tau)\nonumber\\  
     &=& 
 \sum_{n_A} \left[1+ \Phi(z,w,\tau)\right]^{n_A-1}n_A    \sum_{n'_{A}}  P_{\rm binom}(n'_{A};\{n_0,p_A(t-\tau)\}) P_{\rm Poisson}(n_A-n'_A; \mu_A(t-\tau))\nonumber\\
        &=& 
     \sum_{n'_{A}} \sum_{n_A} \left[1+ \Phi(z,w,\tau)\right]^{n_A-1}n_A  P_{\rm binom}(n'_{A};\{n_0,p_A(t-\tau)\}) P_{\rm Poisson}(n_A-n'_A; \mu_A(t-\tau))\nonumber\\
        &=& 
     \sum_{n'_{A}} \sum_{n_A} \left[1+ \Phi(z,w,\tau)\right]^{n_A+n'_A-1}(n_A + n'_A) P_{\rm binom}(n'_{A};\{n_0,p_A(t-\tau)\}) P_{\rm Poisson}(n_A; \mu_A(t-\tau))\nonumber\\
    &=& 
     \sum_{n'_A}  \left[1+ \Phi(z,w,\tau)\right]^{n'_A}\frac{n_0! p_A(t-\tau)^{n'_A} }{n'_A!(n_0-n'_A)!} (1- p_A(t-\tau))^{n_0-n'_A} \nonumber\\
     &\times& \sum_{n_A} \left[1+ \Phi(z,w,\tau)\right]^{n_A-1} e^{-\mu_A(t-\tau)}\frac{\mu_A(t-\tau)^{n_A}}{(n_A-1)!} \nonumber\\
     &+& 
     \sum_{n'_{A}}  \left[1+ \Phi(z,w,\tau)\right]^{n'_A-1}\frac{n_0! p_A(t-\tau)^{n'_A} }{(n'_A-1)!(n_0-n'_A)!} (1- p_A(t-\tau))^{n_0-n'_A} \nonumber\\
    &\times&  \sum_{n_A}\left[1+ \Phi(z,w,\tau)\right]^{n_A} e^{-\mu_A(t-\tau)}\frac{\mu_A(t-\tau)^{n_A}}{n_A!}\nonumber\\
    &=& 
      \left[1+ \Phi(z,w,\tau) p_A(t-\tau) \right]^{n_0}\mu_A(t-\tau)  \exp\left[\mu_A(t-\tau)  \Phi(z,w,\tau) \right] \nonumber\\
    &+& 
    n_0 p_A(t-\tau) \left[1+ \Phi(z,w,\tau)p_A(t-\tau) \right]^{n_0-1}  \exp\left[\mu_A(t-\tau)  \Phi(z,w,\tau) \right] \nonumber\\
    &=& 
      \left[1+ \Phi(z,w,t)  \right]^{n_0}\mu_A(t-\tau)  \exp\left[\mu_A(t-\tau)  \Phi(z,w,\tau) \right] \nonumber\\
    &+& 
    n_0 p_A(t-\tau) \left[1+ \Phi(z,w,t) \right]^{n_0-1}  \exp\left[\mu_A(t-\tau)  \Phi(z,w,\tau) \right] \label{factor}
\end{eqnarray} 
where in order to get the last line, we used the fact that
\begin{equation}
p_A(\tau)p_A(t-\tau) = p_A(t),\quad  p_I(\tau)p_A(t-\tau) = p_I(t)    
\end{equation}
 so that
\begin{eqnarray}
\Phi(z,w,\tau) p_A(t-\tau) &=& (z-1) p_A(\tau) p_A(t-\tau) + (w-1) p_I(\tau) p_A(t-\tau)\nonumber\\ 
&=& (z-1) p_A(t) + (w-1) p_I(t) = \Phi(t).
\end{eqnarray}

Therefore, the right-hand side of Eq.(\ref{nmm}) becomes
\begin{eqnarray}
&&\left[\gamma (1-z)+ \beta (w-z) \right] \partial_z G + \zeta (1-w)\partial_w G + \alpha(z-1)  G(z,w,t)\nonumber\\
    &+& \beta e^{-\zeta \tau} (1-w) \sum_{n_A} \left[1+ \Phi(z,w,\tau)\right]^{n_A-1}n_A P(n_A,t-\tau) \exp\left[\alpha \int_0^\tau dt'\Phi(z,w,t') \right] \theta(t-\tau) \nonumber\\
     &=& n_0  \left[1 + \Phi(z,w,t) \right]^{n_0-1} \left[\left(-a (z-1) + \beta(w-1) \right) p_A(t) - \zeta (w-1)  p_I (t) \right] e^{\mu_A(t)(z-1)+\mu_I(t)(w-1)}\nonumber\\
    &+&  \left[1 + \Phi(z,w,t) \right]^{n_0}  \left[\left(-a (z-1) + \beta(w-1) \right) \mu_A(t) - \zeta (w-1)  \mu_I(t) + \alpha(z-1) \right] e^{\mu_A(t)(z-1)+\mu_I(t)(w-1)} \nonumber\\
    &-& \beta e^{-\zeta \tau} (w-1) \mu_A(t-\tau)     \left[1+ \Phi(z,w,t)  \right]^{n_0}  \exp\left[\mu_A(t-\tau)  \Phi(z,w,\tau) + \alpha \int_0^\tau dt'\Phi(z,w,t') \right]\theta(t-\tau) \nonumber\\
    &-& \beta e^{-\zeta \tau} (w-1)     n_0  p_A( t-\tau)  \left[1+ \Phi(z,w,t) \right]^{n_0-1}  \exp\left[\mu_A(t-\tau)  \Phi(z,w,\tau) + \alpha \int_0^\tau dt'\Phi(z,w,t') \right]  \theta(t-\tau)\nonumber\\
    &=& n_0  \left[1 + \Phi(z,w,t) \right]^{n_0-1} \left[\left(-a (z-1) + \beta(w-1) \right) p_A(t) - \zeta (w-1)  p_I (t) \right] e^{\mu_A(t)(z-1)+\mu_I(t)(w-1)}\nonumber\\
    &+&  \left[1 + \Phi(z,w,t) \right]^{n_0}  \left[\left(-a (z-1) + \beta(w-1) \right) \mu_A(t) - \zeta (w-1)  \mu_I(t) + \alpha(z-1) \right] e^{\mu_A(t)(z-1)+\mu_I(t)(w-1)} \nonumber\\
    &-& \beta e^{-\zeta \tau} (w-1) \mu_A(t-\tau)     \left[1+ \Phi(z,w,t)  \right]^{n_0}    e^{\mu_A(t)(z-1) + \mu_I(t)(w-1)}\theta(t-\tau)\nonumber\\
    &-& \beta e^{-\zeta \tau} (w-1)  \theta(t-\tau)    n_0  p_A( t-\tau)  \left[1+ \Phi(z,w,t) \right]^{n_0-1}  e^{\mu_A(t)(z-1) + \mu_I(t)(w-1)} \theta(t-\tau)\label{rhs}
\end{eqnarray}
where in order to get the last line, I used the fact that
\begin{equation}
    \mu_A(t-\tau) p_A(\tau) = \mu_A(t)-\mu_A(\tau),\quad  \mu_A(t-\tau) p_I(\tau) = \mu_I(t)-\mu_I(\tau), 
\end{equation}
and
\begin{equation}
    \alpha \int_0^\tau \Phi(t') dt' = (z-1) \mu_A (\tau) + (w-1) \mu_I(\tau),
\end{equation}
so that 
\begin{equation}
    \exp\left[\mu_A(t-\tau)  \Phi(z,w,\tau) + \alpha \int_0^\tau dt'\Phi(z,w,t') \right] = e^{\mu_A(t)(z-1) + \mu_I(t)(w-1)} .
\end{equation}

Next, substituting the expression for  $G(z,w,t)$ in Eq.(\ref{genap}) to the left-hand side of Eq.(\ref{nmgen}), we  get
\begin{eqnarray}
    \partial_t G(z,w,t) &=& n_0  \left[1 + \Phi(z,w,t) \right]^{n_0-1}\nonumber\\
    &\times& \left[(z-1) \dot p_A (t) + (w-1) \dot p_I (t) \right] e^{\mu_A(t)(z-1)+\mu_I(t)(w-1)}\nonumber\\
    &+&  \left[1 + \Phi(z,w,t) \right]^{n_0} e^{\mu_A(t)(z-1)+\mu_I(t)(w-1)}\nonumber\\
    &\times& \left[(z-1) \dot \mu_A(t) + (w-1) \dot \mu_I(t)\right]
     \label{lhs}
\end{eqnarray}

Since
\begin{eqnarray}
    \dot p_A(t) &=& -a  p_A(t),\nonumber\\
     \dot p_I(t) &=& \beta p_A(t) -\zeta p_I(t)-\beta e^{-\zeta\tau} p_A(t-\tau) \theta(t-\tau) ,\nonumber\\
    \dot \mu_A(t) &=& \alpha - a\ \mu_A(t),\nonumber\\
    \dot \mu_I(t) &=& -\zeta  \mu_I(t) + \beta \left[\mu_A(t) - e^{-\zeta \tau} \mu (t-\tau)\theta(t-\tau)\right],
\end{eqnarray}
we see that the expressions in Eq.(\ref{rhs}) and Eq.(\ref{lhs}) are equal. Therefore, the distribution in Eq.(\ref{mulpoisol1}) is the solution of the master equation (\ref{nmm}).
\end{document}